\documentclass[referee]{raa}            

\usepackage{graphicx,times}             

\begin{document}

   \title{  Estimation of  the masses of selected stars of Pulkovo program by means of astrometry methods.
}

   \volnopage{Vol.0 (200x) No.0, 000--000}      
   \setcounter{page}{1}          

   \author{N.A.Shakht
      \inst{1}
   \and D.L.Gorshanov
      \inst{1}
   \and O.O.Vasilkova
      \inst{1}
   }

   \institute{Central~(Pulkovo) Astronomical Observatory of Russian Academy of Sciences,
             St-Petersburg,Russia; {\it shakht@gaoran.ru}\\
   }

  \date{Received~~2009 month day; accepted~~2009~~month day}

\abstract{Stars of Pulkovo observatory program are observed on 65-cm refractor~ during many years for study their positions and movement. We represent example of two visual binary stars, for which orbits and masses of components were determined, and two astrometric stars, for which masses of their unseen companions have been estimated. The first stars are: ADS 14636 (61 Cygni) and ADS 7251 and others are: Gliese 623 and~ ADS 8035~(Alpha UMa).
   The direct astrometric methods were used for estimation of mass-ratio and masses.
}

   \authorrunning{N.A.Shakht, D.L.Gorshanov \& O.O.Vasilkova}            
   \titlerunning{Estimation of the masses of  selected stars  }  

   \maketitle

%
%
\section{Introduction}           
\label{sect:intro}

The mass of the star is one of its most important characteristics. Amount of matter in a star determines its temperature and pressure in the center, and also ~determines~ other characteristics~ of the star~ and then its evolutionary path.
Direct estimates of the mass of the star are made on the basis of the law of universal gravitation.The study of binary stars has allowed  to establish the unity of Newton's laws in the Universe and  to obtain the fundamental knowledge about the masses of stars on the basis of observations.
\\
The determination of the masses of the stars by classic way is done by means of positional astrometric observations of double stars with the high-precision instruments~ for~ a long time. Observations should cover a long time interval, or the period of revolution of a double star must be sufficiently short compared to the total observation period. We also need the exact value of the distances for stars and that means - the exact parallax for stars in the solar neighborhood.
Other methods to calculate the mass are considered to be indirect, they are not built on the law of gravity, but on the analysis of the stellar properties (luminosity, temperature, radius) which are somehow connected with the mass.

\newpage
\section{History and observations}
\label{sect:Obs}

At present, more than 100,000  visual double stars are known. WDS and CCDM catalogs respectively have 78 000 and 110 000 such objects and about 3000 calculated orbits. But only for~ some of them, the accuracy of the calculation of the orbits allows us to determine the sum of the masses of the components. Even fewer stars, which you can directly determine the ratio of the masses for.
\
The sum of the masses of the components can be determined from the 3-rd Kepler's law:
\begin{equation}
M_1+M_2=\frac{a^3}{P^2}
\end{equation}
where $a$ is the semimajor axis of the true orbit in a.u. and $P$ $-$ the orbital period in years, $M_1$ and $M_2$ are the component masses in $M_\odot$. ~The sum of masses may be estimate, if the relative orbit of double star and also the distance to it are known.
\\
The masses error~ depends mostly on the uncertainties of the parallax.
\begin{equation}
{\frac{\Delta(M_1+M_2)}{(M_1+M_2)}}    \geq
3 \frac{\Delta\pi}{\pi}
\end{equation}

The ratio of the masses can determine, if the  part of the absolute orbit of each component is known. If you define semimajor axis of the absolute orbits $a_1$ and  $a_2$, we can find the ratio of the masses of the components, as they are inversely proportional to the semimajor axes:
\begin{equation}
 \frac{M_1}{M_2 } = \frac{a_2}{a_1}
\end{equation}

A significant part of the Pulkovo program of photographic observations on a 26-inch (65 cm)refractor was made up for double and multiple stars. Our observations were carried out since 1956 to determine the dynamic parameters, orbits, and masses of visually binary stars.
The list of these stars contains more than 400 objects and about 20,000 astronegatives which belong to Pulkovo glass-library in that there are more then 50 000 plates obtained for more than 100 years of observations.
Since the beginning of the 2000s, observations on 26-inch  refractor have been continued using CCD- cameras.
Most of the Pulkovo programme stars are visually double and represent wide pairs:~~$3''<\rho <30''$
As a result of observations, $\rho$ is a defined mutual angular distance of the components of a binary star AB, and $\theta$ - the positional angle of the direction of AB arc on the celestial sphere.
\\
We used the observations on the long-focus visual refractor D=650 mm,~~ F=10413 mm with a small size of the visible field:
$0.7\times1.0^\circ$. The plates ORWO-NP27 and ORWO-WO1 were used.
During the observations we applied the Hertzsprung's technique of multiple exposures (from 6 to 20 images on the plate),  what was proven in observations of double stars and in the creating parallactical series. In this case, in the narrow field of view on the plates, we usually get a small amount of stars - from 4 to 6 and as shown, see for instance, Izmailov et al. \ (\cite{Izm16}), additional terms, depending on the color and magnitude,cannot be confidently determined.
 Errors occurring due to a difference between colors of the main object and the reference star, associated in particular with atmospheric dispersion, we tried to reduce, making all observations almost strictly in meridian.
\newpage
Photography was also produced with a yellow filter,  installed in the cassette, according to it and  with an appropriate type of plates it was possible to reduce the difference between the spectral sensitivity of the main star and star field. It was  the best way to correspond to 61 Cyg, for that we have possibility to use ortochromatically plates ORWO-WO1.

Magnitude $m_{vis}$  of Gliese 623 equal to $10^{m}.3$, nearly coincides with the average value of the reference stars magnitude , which is equal to $10^{m}.6$.
The brightest star ADS 8035 was observed with additional neutral filter, that attenuates  a magnitude of an object for $6^m$.
 For stars  ADS 7251 and 61 Cyg we have combined series. Part observations were made in the parallactic program with weaking with grating (61 Cyg) and with an gap-loss attenuator,
 but most plates were obtained in the program of double stars without weakening and according to it we have difficulties with selection of plates with sufficient accuracy.

\begin{table}
\begin{center}
\caption[]{ Observational Data
.}
\label{Tab:publ-works}

 \begin{tabular}{clccrl}
  \hline\noalign{\smallskip}
No &  Star  & Observations &N&n& References\\
  \hline\noalign{\smallskip}
1  & Gliese 623B  &1979-1995 &- &90     &
Shakht\ (\cite{shakht95}),(\cite{shakht97})  \\

2  & 61 CygA    &1958-1997   &380&153     & Gorshanov et al.\ (\cite{gor06})                 \\
  &             &1958-2006            & 420&153         &present work\\
3 & 61 CygB       &1958-1997 &380&153     & Gorshanov et al.\ (\cite{gor06})\\
  &             &1958-2006            & 420&153         &present work\\
4  & ADS 7251A   &1962-2005 &  204&115   &  Shakht et al.\ (\cite{shakht10})                 \\
5  & ADS 7251B   &1962-2005 &   204& 115    &Shakht et al.\ (\cite{shakht10})                  \\
6  & ADS 8035B     &1975-2005 & -  &20     & present work                \\
\noalign{\smallskip}\hline
\end{tabular}
\end{center}
\end{table}

Then we represent a little review of results which were obtained during observations with Pulkovo 26-inch refractor.
We have tried to estimate masses of some stars on the basis of positional observations.
Table 1 lists the names of stars, observation intervals, as well as references to work that has already been completed and the results of present work.
 N - the number of plates used to obtain relative positions to determine the orbit and the sum of the masses, n -  the number of plates where the object is measured in the frame of background stars.

At the present time, we summarize the results of photographic observations. Several stars have been chosen for which, in principle, the masses of the components can be determined. However, the study of each object is associated with certain problems. We wanted to test how the accuracy of observations and the ratio of the observed arc to the orbit affect the results.
\\
We had two wide pairs of binaries:~ADS 14636 (61 Cyg) and ADS 7251. And also we had a star with unseen optically component, spectroscopic binary Gliese 623. And our~ new~ attempt is to estimate the mass-ratio of star  ADS 8035 (Alpha UMa).~
 We esimated the mass-ratio, for Gliese 623 - the low limit of mass of the unseen component, also for ADS 8035 the value $\textbf{\textit{B}}-\beta$, where \textit{\textbf{B}} is mass-ratio and  $\beta$ depends from luminosity of the satellite.
\newpage
To determine the orbits of binary stars, we apply the Apparent Motion Parameters (AMP) method developed in Pulkovo, which was repeatedly used to determine the orbits of satellites and asteroids from observations on a relatively short arc of the orbit, see Kiselev et al. (\cite{kis96}).
 We also applied this method to estimating the black hole mass at the center of our galaxy in the system: a black hole + star S01 rotating around the dynamic center,~ see Kiselev et al.\ (\cite{kis07}).
\section{Data processing}
We measured our double stars  to obtain  relative coordinates $\rho$  and $\theta$ ~or~ $\Delta\alpha\cos\delta = \Delta X=\rho$ sin$~ \theta$  and
$\Delta\delta = \Delta Y=\rho$ cos$~ \theta$.
Then we have measured each component with respect to references stars system. We used
the method of six constantes and the reduction to the standard plate. Here we give
the basic formula for calculations. We used the methods outlined in classical works, such as the book of Van de Kamp
(\cite{VdK81}). In common case they are followings for the main component A and also for the component B:
\begin{equation}
~X_{A}  = C_x + \mu_x (t-t_0) + \pi P_x + q_x t^2~- \textbf{\textit{B}}\Delta X~,~~
~~~Y_{A}  = C_y + \mu_y (t-t_0) + \pi P_y + q_y t^2~- \textbf{\textit{B}}\Delta Y~~~~~~~~~~~~~~~~~
\end{equation}
\begin{equation}
~X_{B}  = C_x + \mu_x (t-t_0) + \pi P_x + q_x t^2~+ \textbf{\textit{A}}\Delta X~,~~
~~~Y_{B}  = C_y + \mu_y (t-t_0) + \pi P_y + q_y t^2~+ \textbf{\textit{A}}\Delta Y~~~~~~~~~~~~~~~~~
\end{equation}
where:~~~~
$\textit{\textbf{B}}=M_B/(M_A+M_B)$,~~~~
$\textit{\textbf{A}}=M_A/(M_A+M_A)=1-\textit{\textbf{B}}$~~~~
\\

and others values are known ones:
 $X_{A}, Y_{A}$ , $X_{B}, Y_{B}$ are positions of star on the basis of reference stars
with respect to selected standard plate,  $C_{xy}$ - position of the center of mass with respect to selected component on standard plate, $\mu_{xy}$,~ $\pi$, $P_{xy}$ and $q_{xy}$ are proper motion, parallax, parallactic factors and quadratic term correspondly. Here $q$ is a prospective acceleration equal to
$q ~=~-1''.24\times10^{-6}\mu \pi V_r$ and it is a significant value only for very nearest stars,
      $ \Delta X, \Delta Y$ - relative positions of B component to A  in  the  moment of observations.

Next, we solve the equations   with respect to the constants $C_{xy}$, the proper motion of the center of mass  and the mass-ratio. If the sum of mass is known, we can calculate the mass of each component $M_{A}$ and $M_{B}$.

In some cases the orbital motion, to a high degree of approximation may be represented as a quadratic time effect.
Then the mass-ratio is determined from the quadratic term $Q$ in equations (6) after excluding  perspective acceleration from it.
\begin{equation}
X  = C_x + \mu_x (t-t_0) + \pi P_x + Q_x t^2~,~~~~~~~~~~~~~~Y  = C_y + \mu_y (t-t_0) + \pi P_y + Q_y t^2 ~~~~~~~~~~~~~~~~~~~~~
\end{equation}

In the practic taking into account our geographical location we prefer to exclude
parallactic replacement by means of catalog's parallax and our observations for selected stars
have been made in limited interval of time each year and near to meridian.
As a rule we applied the following equations:
\begin{equation}
X'_A = C_x + \mu_x(t-t_0)  - \textbf{\textit{B}}\Delta X~,~~~~~~~~~~Y'_A = C_y + \mu_y (t-t_0) - \textit{\textbf{B}}\Delta Y~~~~~~~~~~~~~~~~~~
\end{equation}
\begin{equation}
X'_B = C_x + \mu_x(t-t_0)  + \textbf{\textit{A}}\Delta X~,~~~~~~~~~~Y'_B = C_y + \mu_y (t-t_0) + \textit{\textbf{A}}\Delta Y~~~~~~~~~~~~~~~~~~
\end{equation}
\newpage
where $X'_A$, $X'_B$, $Y'_A$, $Y'_B$ are corrected for parallax and accelration.
\section{Data analysis}
Now we would like to give some examples and to note some specific
problems with treating for each case.
\\
I. At first we consider two stars-components of binary:

61CygA  [ADS~14636A, GJ820A, HIP104214, HD201091,~$5^m.21$,~K5V,
~~$21^h06^m .9$,~	$+38^o 45'$, ~$\pi=0''.286$,~	$Vr=-64.7~$ km/s].
\\
61CygB  [ADS~14636B, GJ820B, HIP104217, HD201092,~$6^m.03$,~K7V,	
~~$21^h06^m.9$,~$+38^o 44'$,~$\pi=0''.286$,~   $Vr= -65.7$ ~km/s].
\\
 For this star, we have their relative positions B-A,
obtained in 1958-2006 with error for normal point 0".007 in $\rho$ and $0^\circ.02$  in $\theta$.
 Positions of each component on the basis of
reference stars on the plates were obtained on interval 1958-1997. Earlier we have the estimate of orbital elements of this pair, see Gorshanov et al.\ (\cite{gor06})
Now we reexaminate our results with additional series of observations 1997-2006.
 Two variants of orbits, obtained with using of parallax $0''.2861\pm0''.0005$ from RECONS.org web-site, are given in Shakht et al.\ (\cite{shakht17}).~We have two estimates of the sum of the masses with control on O$-$C. Only on our 1958-2006 observations,
 we have estimated the sum of masses as $1.31 M_\odot$ and with jointing of all observations from WDS as
$1.4M_\odot$.

At first, we applied the classic formulas to the stars studied. The  values of  mass-ratio
for 61 Cyg are in table 2. The results of the decision on the two  coordinates are given for  components A and B. Average error unit of weight $E_0$ is about $0''.044$. In table 2 in columns 2-4 are the results for masses of  components.~~Perhaps the observational period of 40 years is not enough for a confident decision  due to masses A and B have significant differences.
\begin{table}
\begin{center}
\caption[]{Mass-ratio from solving for A and B components 61 Cyg
with projection on X,Y coordinate
.}\label{Tab:publ-works}
 \begin{tabular}{ccccc}
  \hline\noalign{\smallskip}
Component& $A_x$ & $A_y$ & $B_x$&$B_y$        \\
\noalign{\smallskip}\hline
 $\textit{\textbf{B}}$    & $0.38\pm0.04$  &$0.43\pm0.06$ & $0.46\pm0.06$& $0.44\pm0.10$
  \\
 $E_0$ & 0".037 & 0".041 & 0".057 & 0".040\\
\noalign{\smallskip}\hline
\end{tabular}
\end{center}
\end{table}

\begin{table}
\begin{center}
\caption[]{Sum and mass-ratio 61 Cyg
.}\label{Tab:publ-works}

 \begin{tabular}{ccccccc}
  \hline\noalign{\smallskip}
$\Sigma M/M_\odot$ & $\textit{\textbf{B}}$  from $A_{xy}$ & $M_A/M_\odot$ &$M_B/M_\odot$& $\textit{\textbf{B}}$  from $B_{xy}$& $M_A/M_\odot$ &$M_B/M_\odot$  \\
\noalign{\smallskip}\hline
1.20 &0.40 &0.72  &0.48 &0.45 &0.66 &0.54\\
1.31 &0.40 &0.79  &0.52 &0.45 &0.72 &0.59\\
1.40 &0.40 &0.84  &0.56 &0.45 &0.77 &0.63\\
\noalign{\smallskip}\hline
\end{tabular}
\end{center}
\end{table}

\begin{table}
\begin{center}
\caption[]{ Estimations of mass-ratio and stellar masses
.}\label{Tab:publ-works}

 \begin{tabular}{cccrl}
  \hline\noalign{\smallskip}
61 Cyg&&&&\\
  \hline\noalign{\smallskip}
$\Sigma M$ & $\textbf{\textit{B}}$   & $M_A$ &$M_B$& References\\
\noalign{\smallskip}\hline

  -              &   0.38       & -&-        &Van de Kamp\ (\cite{VdK40})\\
  -              &   0.55        & -&-       &Van de Kamp\ (\cite{VdK53})\\
  -              &   0.47       & -&-        &Van de Kamp\ (\cite{VdK53})\\
 1.12      &0.52 &0.54&0.58                  &Van de Kamp\ (\cite{VdK54})\\
 1.26              &0.47      &0.67 &0.59    &Walker et al.\ (\cite{Walk95})\\
 1.30   &0.46 &0.69  &0.60                   &Kervela et al.\ (\cite{Kv08})\\
 1.31   &0.48 &0.68    &0.63                 &Boyajian et al.\ (\cite{Boy12})\\
 1.31   &0.40 &0.79    &0.52                 &Present work \\
 1.31   &0.45 &0.72    &0.59                 &Present work \\
 1.0    &0.5  & 0.5    &0.5                  &Cester et al.\ (\cite{Cest88})\\
\noalign{\smallskip}\hline
ADS 7251&&&&\\
\noalign{\smallskip}\hline
$\Sigma M$ & $\textbf{\textit{B}}$   &$ M_A$ &$M_B$& References\\
\noalign{\smallskip}\hline

 0.91    &0.505   &0.45&0.46     &Hopmann\ (\cite{Hop54})                 \\
 2.26    &0.50   &1.13&1.13     &G\"{u}ntzel-Lingner\ (\cite{Gun55})\\
 1.14   &0.64   &0.41&0.73    &Chang\ (\cite{Chang72})\\
  1.22   &0.49 &0.62    &0.60                 &Boyajian et al.\ (\cite{Boy12})\\
  1.10&0.49&0.56&0.54&present work\\
 1.10&0.48&0.57&0.53&present work\\
 \noalign{\smallskip}\hline
\end{tabular}
\end{center}
\end{table}
\newpage
In addition, we managed to solve equation with respect to  quadratic terms $Q_x$ for A and B components  which turned out to be equal to
 $ -0.00008 \pm 0.00002 ''/yr^2$ and $0.000205 \pm 0.000007''/yr^2$ for A and B, respectively.
The perspective acceleration for this star, which equaled $0.000079''/yr^2$, was subtracted from $Q_x$ and corrected values for A and B coordinates $Q'_a$ and $Q'_b$ were obtained.
As a result, the value $M_b/M_a =- Q'_a/Q'_b$ equaled 0.78 and mass-ratio \textbf{\textit{B}} equaled 0.43 were obtained.~ In table 3, ~the masses of components 61 Cyg are given for three values of the sum of masses and for two
mass-ratio obtained with different ways.
As noted by Kervella et al.(\cite{Kv08}), the dynamic mass of this star is determined uncertainly and it is associated with difficulties due to a large period of the orbit  about $6\div 7$ centuries.
\\
The determination of the component masses for the sum of masses $1.31M_\odot$ and the ratio of 0.43 is most satisfactory in our results, although there is remarkable difference between the masses.
Now we limited the choice of materials by errors of the unit weight $E_0$ in the range $0''.018\div0''.045$ and in the interval 1958-1997.At present, we hope to precise this result by adding data for 1997-2006, that can provide a greater difference of epochs and also weight  of parameters.
\\
  Only mass-ratio for 61 Cyg was presented in earlier work due to absence of  reliable orbits and sum of masses. Our results obtained of the solving for component B  with $\textbf{\textit{B}}$
are more according to spectral classes components than for component A.
\newpage
II. Then we had analyzed the astrometric and spectral stars Gliese 623 which was observed as a single star at Pulkovo in $1979-1995$ years.

[GJ623, ~HIP 80346,~ AC$+48^o 1595/89,~
 10.^m3,~ M3.0Ve,~ 16^h 24^m.1,~ +48^o 21',~
\pi= 0.'' 124$, ~$ V_r~=~-26.3$ km/s].
Here we had the opportunity to explore the residuals~ $R_x$, $R_y$ obtained in the movement of the star after elimination of proper motion and parallax.
$P,~To$ and $e$ were founded by means of graphical method. Then using the values $P,~To$ and $e$ the coordinates
of photocenter's ellipse in units of semimajor axis $x,y$ were calculated.
We solved (4)  for ~$ C_{xy}$,~ $\mu_{xy}$,~ $\pi$ and then  considered residuals:~ $R_x, R_y$.~Residuals characterize the orbital motion of the photocenter of the system:
\begin{equation}
R_x = C_x + x(B) + y(G),~~~~~~    R_y = C_y + x(A) + y(F),
\end{equation}
where $x,y$ are the rectangular coordinates of the ellipse in units of the semimajor axis. ~(B), (G), (A), (F) are constants of Thiele-Innes, that contain the information about the orientation of the orbit of the photocenter ~and its semimajor~ axis~ $a_1$,~ which is the size of the astrometric signal.~ We determined the mass of the satellite,  according to the formula:
\begin{equation}
M_2=\frac{a_1^"}{\pi}(\frac{M_1+M_2}{P})^{2/3}
\end{equation}
\vskip 2 mm

by the method of successive approximations, assuming the mass of the principal star equal to $0.31M_\odot$, corresponding to its spectral class, and taking in the first approximation the mass of the component equal to zero.
\\
Various authors, Lippincott \& Borgman (\cite{Lip78}), Marcy \& Moore (\cite{Marc89}),  have estimated the mass of the invisible satellite in the range $0.06\div0.08M_\odot$ and $0.067\div0.087 M_\odot$.
 We estimated the lower mass limit of the satellite as $0.09 \pm 0.03 M_\odot$, taking into account the obtained trigonometric parallax $0''.131$, ~Shakht~(\cite{shakht95}), Shakht~ (\cite{shakht97})~~and its catalog value $0''.124$.
\\
Astrometric observations, which were according to McCarthy\& Henry~(\cite{Mc87}) spectral observations, and our parallax,  different from photometric one ($0".151$), showed that the satellite is a stellar object, what was confirmed later.
In 1996, the Gliese 623 star was resolved into two components using the Hubble telescope, Barbieri et al.~(\cite{barb96}). According to observations with modern technics with a 200-inch Palomar telescope with adaptive optics, see Martinache et al.~(\cite{Mart07}) and with a telescope Keck II in the NIR-range, the satellite's mass  is found to be $0.115 \pm 0.002 M_\odot$. At present the improved value is appeared $0.114\pm 0.002M_\odot$,  Benedict et al.~(\cite{Ben16}).
We cannot reach such accuracy now,  but we   continue CCD- observations to detect any long-period effects in the motion of these stars.\\
\newpage
III. Dubhe (Alpha~UMa) [ADS8035, BU1077, HD95689, HIP54061,~
 $1^m.79$,~  G9III,~$ 11^h 03^m.7,~ +61^o 45',~
\pi= 0''.026, V_r  =-9.4 $ km/s] is multiple star and spectroscopic binary ADS 8035 with optically unseen component B and with remote component C.
\\
Photographic observations of ADS 8035A with 26-inch refractor were made in 1975-2005. It was observed as single one because   a limit of separation of  26-inch refractor is $1''.5$  and so  the   satellite is not   observed.
But we can trace the path of the main star on the background of reference stars.
\\
The period of revolution of the satellite is 44.4 years, therefore it is possible to determine the relative astrometric orbit of the photocenter or, taking the orbital elements to be known, to estimate the mass-ratio possibly with greater accuracy than on observation of a short arc of the relative orbit of visual binary.
For ADS 8035  we calculated the ephemeris using the relative orbit of the satellite on elements, given in Scardia et al.,~(\cite{Sca11}). In this way, relative positions $\Delta {Xj}$, $\Delta {Yj}$ corresponding to our moments of observations were obtained.
\\
The course over time of relative positions of the visible components A is shown on Fig.1. For this star we have a little number plates  and we selected 20 test plates.  Then the following equations (10)solved , for which parallactic displacement was excluded previously and so only C and $\mu$ were determinated.
\\
Residuals $R_x$, $R_y$ remained from this solution and consisted orbital motion  $\textbf{\textit{B}}\Delta X_j$ and   $\textbf{\textit{B}}\Delta Y_j$ ~  depending on time~ (1900+)~   are given in the figure 2 and 3.
\begin{equation}
~X_j = C_x + \mu_x(t_j-t_0)  - \textbf{\textit{B}}\Delta X_j~,~~~~~~~~
~~~~~~Y_j = C_y + \mu_y (t_j-t_0)   - \textbf{\textit{B}}\Delta Y_j~~~~~~~~~~~~~~~
\end{equation}
 As a result after solving of equations (10), the following parameters were obtained:
$C_x= -0.''230 \pm 0''.012$,~~         $\mu_ x = -0''.1240 \pm 0''.0007$/yr,~~
$C_y=  0''.008 \pm 0''.016$,~~        $\mu_ y = -0.''0625 \pm 0''.0009$/yr.~~
\\
The value of  $\emph{\textbf{B}}$~~ turned out to be equal
     $0.324\pm 0.024$~~as a result of the solution with respect to the X-coordinate.
Error of the unit of weight for X- coordinate $E_0$ is equal to  $0''.038$.
\\
The accuracy for Y-coordinate is worth than for X, the value of
$\textbf{\textit {B}}$~
is obtained with a  lower weight and mass-relation $\textbf{\textit{B}}$~~is equal $0.34\pm 0.13$.
\\
We did many different variants of solving  suggesting, that mass-ratio may be less. But we applied our variant depending   to  minimal $E_0$ $–$ error of unit weight (one equation).
$M_A$ and $M_B$ can be estimated as 3.9$M_\odot$ and 1.8~$M_\odot$.
Here the  mass-ratio corresponds to its lower limit  $\textbf{\textit{B}}_{lower}=\textbf{\textit{B}}-\beta$,
due to the fact that the satellite is located close to the main star and has its own luminosity.
\newpage
\begin{figure}
   \centering
   \includegraphics[width=98 mm, height=48mm]{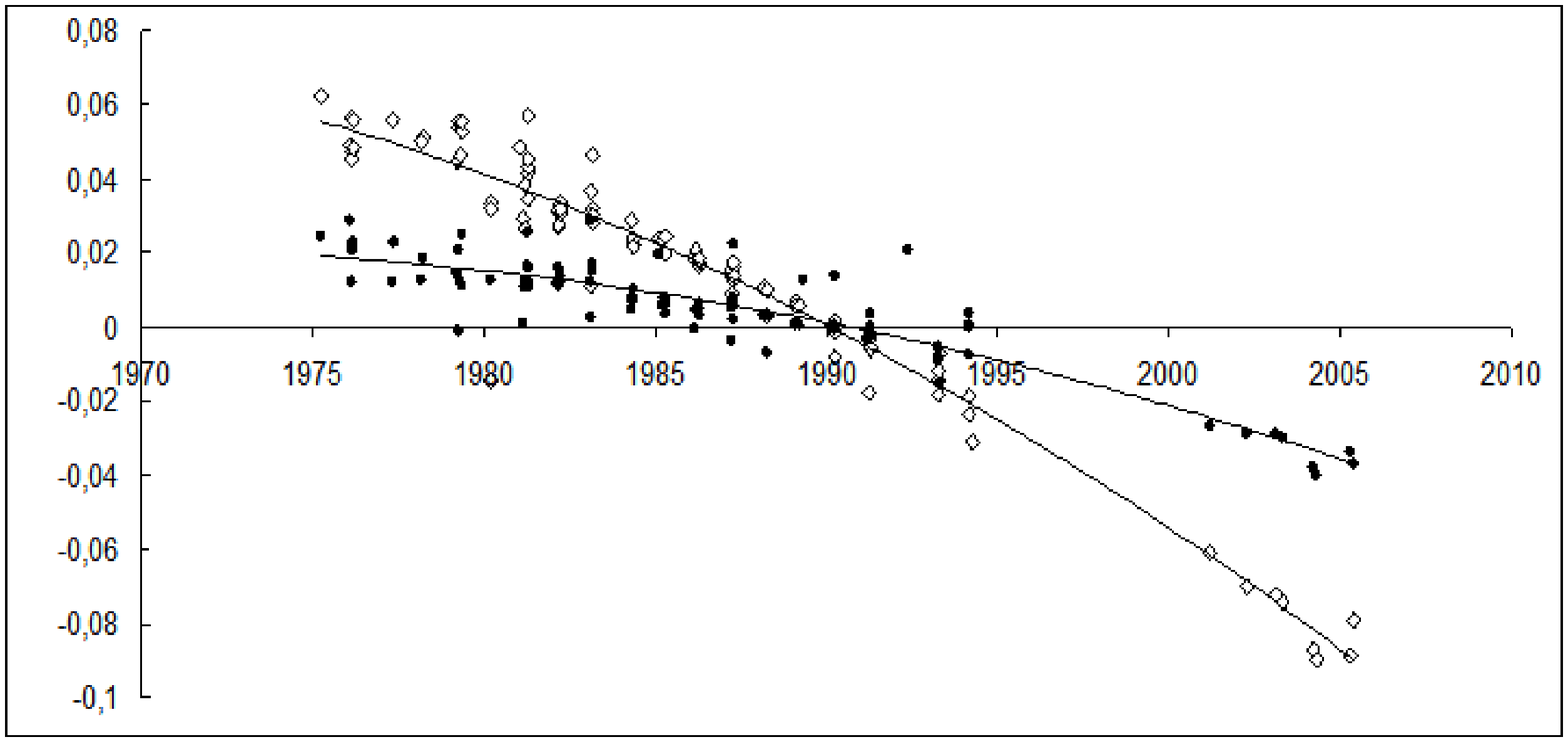}
   \caption{ The course with the time of coordinates $X_j$,$Y_j$ of ADS 8035A  in mas}
   \label{Fig1}
   \end{figure}
\begin{figure}[h]
  \begin{minipage}[t]{0.495\linewidth}
  \centering
   \includegraphics[width=40mm,height=42mm]{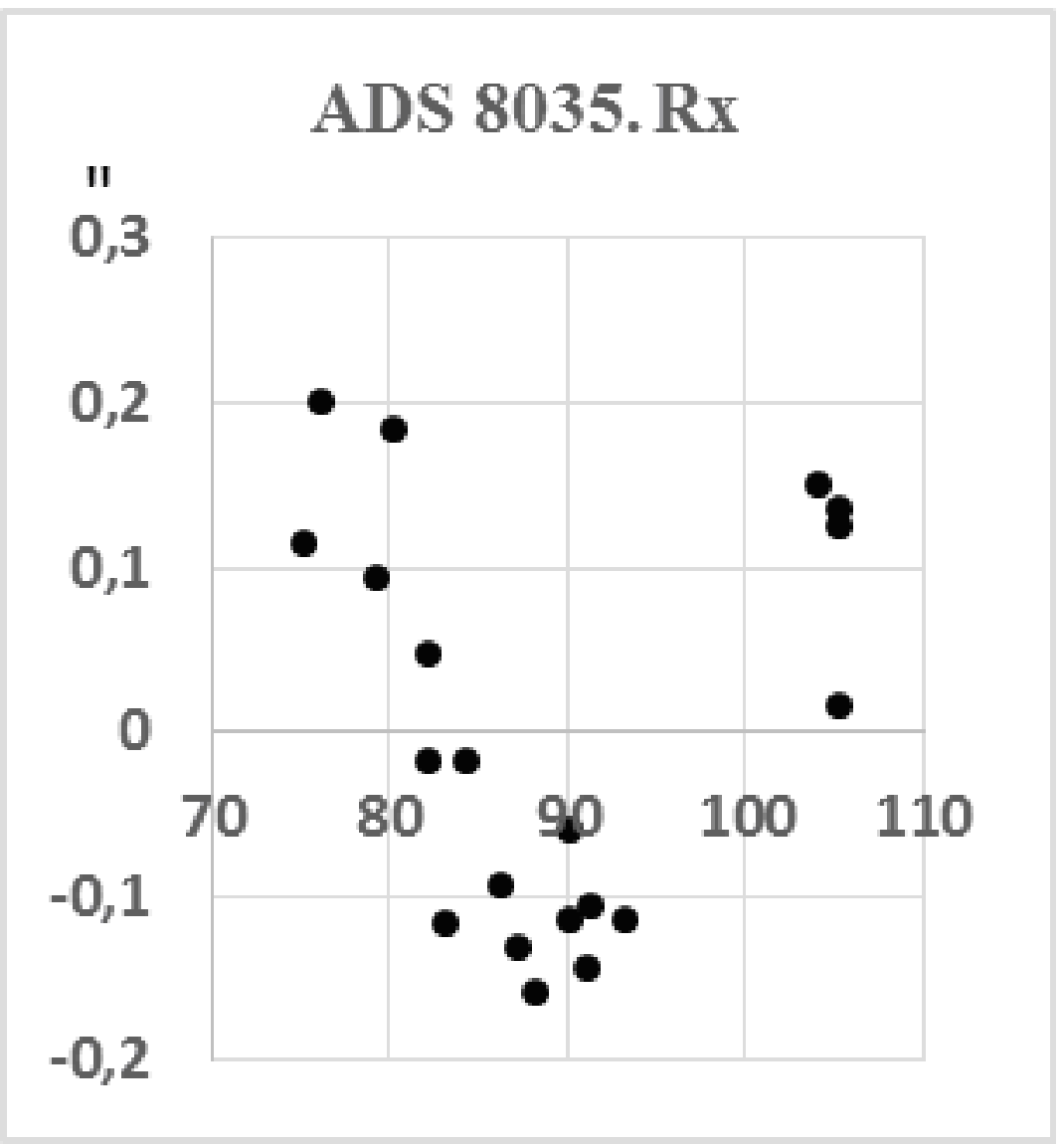}
   \caption{{\small Residuals $R_x$ }}
  \end{minipage}%
  \begin{minipage}[t]{0.495\textwidth}
  \centering
   \includegraphics[width=40mm,height=42mm]{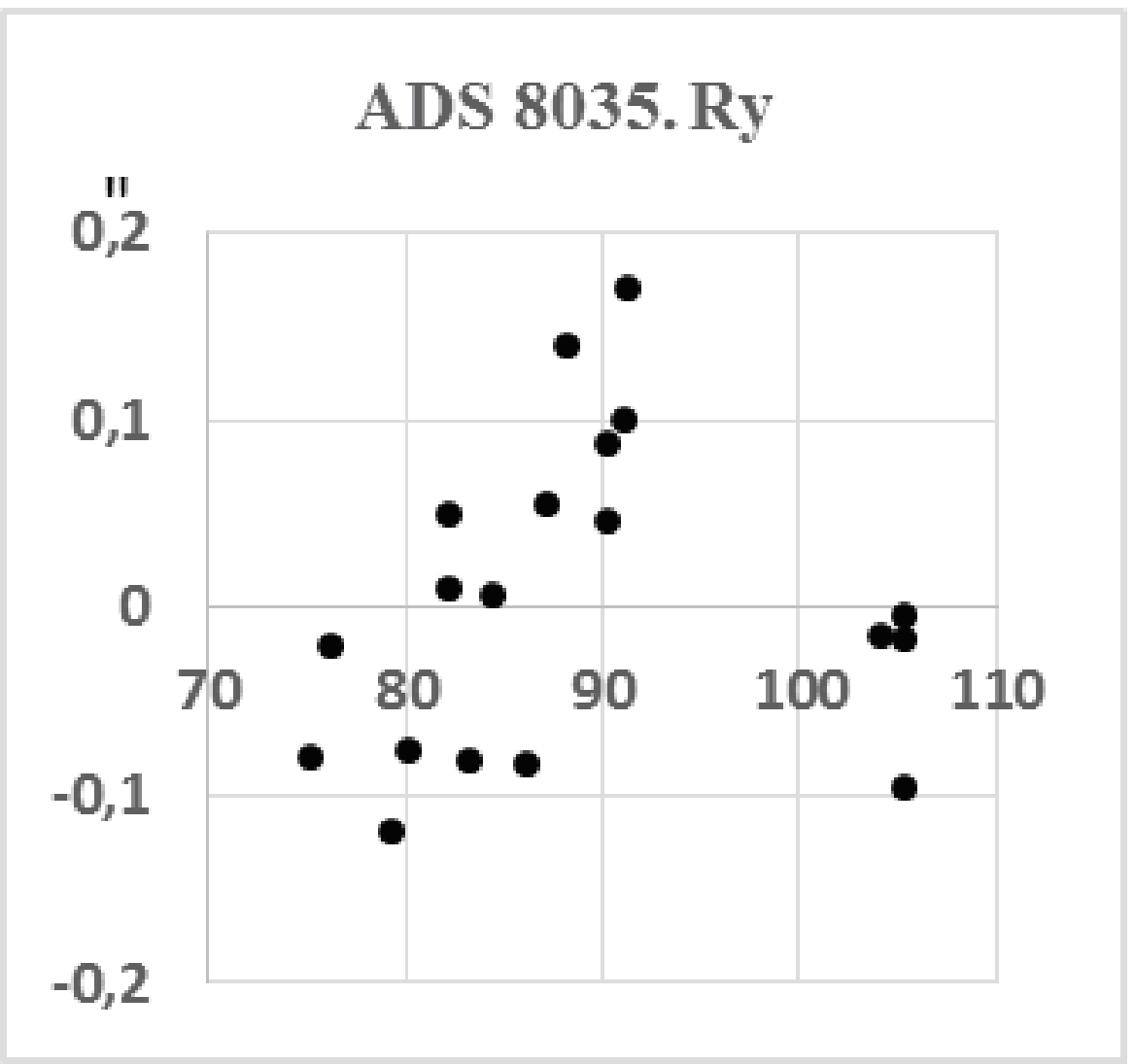}
  \caption{{\small Residuals $R_y$ }}
  \end{minipage}%
  \label{Fig:fig23}
  \end{figure}
\newpage
IV. ADS 7251A  [GJ338A,~ HIP45343,~ HD79210,~
$7^m.6$,~   K7V,~ $09^h 14^m.4,~ +52^o 41'$ ,~
$\pi=	0''.164,~ V_r$ = 11.1 km/s].
\\
~~~ADS 7251B~~ [GJ338B,~~ HIP120005,~ HD79211,~
$7^m.7$,~   M0V,~ ~$09^h 14^m.4, +52^o 41'$,~
$\pi=	0''.164,~ V_r$ = 12.5  ~km/s~].

Another example of obtaining the sum and the mass-ratio is a calculation of the corresponding values after the determination of the relative orbit and the sum of the masses for the visual binary star ADS 7251 on the basis of observations with the Pulkovo 26-inch refractor.
The results were obtained in  1962 -2006, with errors of the  distance $\rho$ and position angle $\theta$  equaled to $0''.006$ and  $0.^\circ02$, respectively.
We have the star with long-period orbit  more than 1500 years.

 Determination of orbit and the estimation of mass were made at Pulkovo,see Shakht et al. (\cite{shakht10}), however observational arc is enough short.
In the solving classical equations (5), there were correlations between unknowns and they hardly separated.
\\
For our case, we applied the  method in which the mass ratio of $\textbf{\textit{A}}= M_{A}/(M_{A}+M_{B})$ and $\textbf{\textit{B}}= M_{B}/(M_{A}+M_{B})$  were considered as the free parameters that are linked together in such a way that $\textbf{\textit{A}}+\textbf{\textit{B}}=1$ and we selected them trying to get the minimum  of the error unit of weight  $E_0$.

The equations  were solved, where $X_{A_{j}}$, $X_{B_{j}}$, $Y_{A_{j}}$, $Y_{B_{j}}$ –positions of components A and B relative to the reference stars for each of the j-th moment, corrected for parallax and acceleration
and the  $C_{xy}$  are the positions of the center of masses with respect to the selected zero-point on the standard plate, $\mu_{xy}$ - proper motion of the center of masses of the system.
\begin{equation}
X_{A_{j}}\textbf{\textit{A}}+X_{B_{j}}\textbf{\textit{B}} = C_{x} + \mu_{x} (t_{j}-t_0), ~~~Y_{A_{j}}\textbf{\textit{A}}+Y_{B_{j}}\textbf{\textit{B}} = C_{y} + \mu_{y} (t_{j}-t_0)
\end{equation}
\begin{equation}
~~~H_{A_{j}}\textbf{\textit{A}}+H_{B_{j}}\textbf{\textit{B}} = C_{h} + \mu_{h} (t_{j}-t_0)
\end{equation}
\newpage
Then we decided to improve the estimate of the mass -ratio using a projection onto a coordinate axis where the correlation with proper motion should not be significant. A system of equations was then solved where the initial relative positions A and B were calculated in projection onto an axis H perpendicular to the direction of motion of the center of masses of the system.
\begin{equation}
H_{A} = -X_{A}\cos\chi + Y_{A}\sin\chi,~~~~ H_{B} = -X_{B}\cos\chi + Y_{B}\sin\chi,
\end{equation}
where $\chi=\arctan(\mu_{x}/\mu_{y})$.  Here the proper motion of center of mass is not determined, while $\mu_{h}$ represent the measurement errors and $C_{h}$ is the summed proper motion of  reference stars projected onto the axis H.
Finally with the mass-ratio~  $\textbf{\textit {B}}$ value of 0.48
we obtained a minimum of $E_0$ equal to $0''.032$ and the mass of the components of 0.57 and 0.53$M_\odot$ that is satisfactory  to spectral classes of components.
In the table  the sum of mass is given according to our estimate, see Shakht et al. (\cite{shakht10}), n-the number of used plates, X,Y - the average result from X,Y axis projections.
\\
\begin{table}
\begin{center}
\caption[]{Sum and mass-ratio ADS 7251
.}\label{Tab:publ-works}

 \begin{tabular}{ccccccc}
  \hline\noalign{\smallskip}
$\Sigma M/M_\odot$ & $\textit{\textbf{B}}$   & $M_A/M_\odot$ &$M_B/M_\odot$& $n$  & $E_0$ & Axis  \\
\noalign{\smallskip}\hline
1.1 &0.46 &0.59  &0.51 &146 &0.070 &X\\
1.1 &0.49 &0.56  &0.54 &104 &0.036 &X,Y\\
1.1 &0.48 &0.57  &0.53 &115 &0.032 &H\\
\noalign{\smallskip}\hline
\end{tabular}
\end{center}
\end{table}
With $E_0=0".032$ we obtained mass-ratio $0.48\pm 0.06$, the errors of masses is not less than $\pm0.1M_\odot$.

V. On Pulkovo observations of homogeneous long-term series, we can estimate the minimum sum of masses for binary stars rotating in elliptical orbits, using their  parameters of motion, even in the event that the orbit itself is not defined.
\begin{equation}
M_1+M_2 > \frac{\rho V^2}{8 \pi^2 \pi_{tr}}
\end{equation}
\begin{equation}
M_1+M_2\geq \frac{\rho \mu}{4 \pi^2 \rho_c \pi^3_{tr} |\sin(\psi- \theta)|}
\end{equation}
\\
In formulas (15-16), $V$ is the  space velocity, $\mu$ and $\psi$ -- relative proper motion of components and its positional angle and $\rho_c$ is the radius of curvature of orbit.
 In some cases, an excess of mass was detected. In this way, the presence of a third companion
 was confirmed for some stars, such as for example,  ADS 497 and ADS 11061,~see, ~Kiselev\&Romanenko~ (\cite{kis96}),
~Kiselev\&Kiyaeva~ (\cite{kis04}),~Tokovinin\&Smekhov~ (\cite{Tok02}).
\newpage
\section{Conclusions}
We have shown several examples of determining the mass using astrometric positional observations. We will make a few remarks on the processing.
\\
1.It should be noted that the sought values were obtained within the precision of positions of the reference stars.
\\
2. In some cases, the mass-ratio can be detected only by one coordinate (RA or Decl.), since on the~ other~ coordinate was obtained with a low weight.
\\
3. If the observed arc is small with respect to the orbit, then a correlations between unknowns can appear in result of solving equations (4), and the required parameters are difficult to separate.
\\
4.There are the cases where the direction of movement of  center of system is close to the direction of orbital motion.
In this case, it is preferable to solve the corresponding equations with projection on the direction of motion or perpendicular to it in order to exclude its proper motion in advance or to solve the equations with respect to the constant and the mass-ratio only.And such approach has been applied for ADS 7251 solution.
\\
5. Because of our geographic location, we could not always receive parallax with the required accuracy, so the parallactic effect,  with the using its catalogue value, was previously excluded from the initial conditional equations.

\section{Acknowledgment}
We wanted to present a part of our studies on the long-time positional observations on Pulkovo 26-inch
refractor and we thank all the observers, who created this base for such research.
\\
We thank also the editorial board of the Journal "Research in Astronomy and Astrophysics"
for the possibility of publishing this article.

\end{document}